# TOF-PET DETECTOR CONCEPT BASED ON ORGANIC SCINTILLATORS

P. Moskal, T. Bednarski, P. Białas, M. Ciszewska, E. Czerwiński, A. Heczko, M. Kajetanowicz, Ł. Kapłon, A. Kochanowski, G. Konopka-Cupiał, G. Korcyl, W. Krzemień, K. Łojek, J. Majewski, W. Migdał, M. Molenda, Sz. Niedźwiecki, M. Pałka, Z. Rudy, P. Salabura, M. Silarski, A. Słomski, J. Smyrski, J. Zdebik, M. Zieliński.

Jagiellonian University, Cracow, Poland

**Abstract:**
In this contribution we present a new concept of the large acceptance detector systems based on organic scintillators which may allow for simultaneous diagnostic of large fraction of the human body. Novelty of the concept lies in employing large blocks of polymer scintillators instead of crystals as detectors of annihilation quanta, and in using predominantly the timing of signals instead of their amplitudes.

**Introduction:**
Positron Emission Tomography is at present one of the most technologically advanced diagnostic method that allows for non-invasive imaging of physiological processes occurring in the body. It plays a fundamental and unique role both in medical diagnostics and in monitoring effects of therapy in particular in oncology, cardiology, neurology, psychiatry and gastrology. However, PET scanners are very expensive: average cost of the device with the necessary software, equipment and adaptation in the room amounts to about four million Euro. Therefore, in Poland there are only twelve such facilities, which is (as per the number of residents) twenty two times less than in the USA where there are currently over two thousand PET/CT scanners [1]. Scanners purchased in Poland are largely the result of a dedicated government program called "National Programme for Fighting Cancer" [2] realized in recent years and it is utterly unrealistic for polish healthcare to reach the level of the USA in the next decades without a technological breakthrough which would allow for a drastic reduction in production cost of PET scanners. In this article we first describe briefly detection technique used in current PET modalities and then we present a new concept of TOF-PET detectors based on polymer scintillators promising significant reduction of costs of PET scanners.

**Current Solutions:**
Currently all commercial PET devices use inorganic scintillator materials as radiation detectors (usually these are the BGO (GE Healthcare), LSO (Siemens) or LYSO (Philips) crystals [3]).
Gamma quanta hitting the scintillator may transfer a part or all of their energy to electron, which, in turn, in the processes of ionization and excitation of atoms or molecules, induces the light flashes, which are then converted into electrical signals by photomultiplier tubes connected to the scintillator. Scintillation crystals, usually blocks with dimensions of approximately 5cm x 5cm and a thickness of about 2.5 cm, are additionally divided into smaller components with dimensions of about 5 mm x 5 mm. At the rear of each block four photomultipliers convert the light pulses into electrical signals. The amplitude distribution of these pulses allows determination of a place where the gamma quantum reacted with accuracy equal to the size of a small crystal element. In further analysis, in order to determine line along which the annihilation quanta were emitted (line-of-response) it is assumed that the gamma quantum has been absorbed in the middle of the detector element. This assumption is one of the essential contribution limiting resolution of the images. All current scanners use for the detection of gamma quanta the photoelectric effect and in the event selection the energy window, typically in the range from 350 keV to 650 keV, is applied [4]. This reduces the noise caused by the annihilation quanta scattering in the patient's body. Such energy window corresponds to the angular range of scattering from 0 to about 60 degrees.

A way to improve the resolution of the tomographic image is determination of the annihilation point along the line-of-response based on measurements of the time difference between the arrival of the gamma quanta to the detectors. This technique is known as TOF (time of flight), and tomographs which use the time measurements are termed TOF-PET. In practice, due to the finite resolution of the time measurement, it is possible to determine only a section along LOR in which the annihilation had occurred with the probability density determined by the time resolution. This improves the reconstruction of PET images by improving signal to noise ratio due to the reduction of noise propagation along the LOR during the reconstruction [5]. Attempts to use the time of flight in PET tomography have been undertaken since 1980 [6]. However, due to the use of inorganic scintillators characterized by very slow signals one could not obtain significant progress. Only the discovery of new LYSO and LSO crystals allowed for more efficient use of time measurements, but they still give signals of significantly longer decay and rise times than organic scintillators. A first commercial TOF-PET constructed in 2006 by PHILIPS based on LYSO scintillator crystal achieved 650 ps (FWHM) for time-of-flight resolution. In 2008 a prototype made by SIEMENS achieved the time resolution of about 550 ps with a scanner based on LSO crystals, which corresponds to the spatial resolution along the line-of-response amounting to 8 cm. Presently, there is ongoing research aiming at discovery of new crystals, which would have better timing properties [6,7].

As regards the scanning area the newest PET scanners manufactured by Siemens and Philips, are built from 24 to 29 thousand crystals combined with over 400 photomultipliers. Detector rings have an internal diameter ranging from 70 cm to 90 cm, width of about 18 cm and allow for imaging of the patient's body over a length of 190 cm. However, the image of the whole body is made in several steps taken moving the patient many times inside the tomograph by about half the width of the ring [5,3]. At present, in order to get an image of the patient lengthwise over 180 cm one requires approximately 17 independent scans. With the current solutions the increase of the rings width in order to cover the entire human body is economically unrealistic, because the number of crystals, photomultipliers and electronics modules increase linearly with increasing longitudinal field of view.

**New Concept:**
In order to decrease the costs of the PET scanners and to improve its TOF resolution we strive to build a TOF-PET detector using polymer scintillators instead of inorganic crystals. Novelty of the concept lies in employing predominantly the timing of signals instead of their amplitudes. The solution proposed will allow for the determination of position and time of the reaction of the gamma quanta based predominantly on the time measurement.

The plastic scintillators were so far not considered as potential sensors for PET detector due to their low density (1.03 g/cm$^3$) and small atomic number of elements constituting the material. Fast organic scintillators are composed mainly of carbon and hydrogen. Small atomic number corresponds to small probability that gamma quanta transfer all their energy to the electrons in the scintillator through the photoelectric effect. Moreover, small density implies a small efficiency for the detection of gamma quanta. Linear absorption coefficient of gamma quanta with an energy of 511 keV for the plastic scintillator equals to 0.098 cm$^{-1}$ [8], and is more than eight times smaller than for the LSO crystals amounting to 0.821 cm$^{-1}$ [6]. Therefore, the efficiency of the reaction of annihilation quantum in the scintillator layer with a thickness of 2.5 cm is 0.217 for the plastic and 0.872 for the LSO crystal. Consequently, the probability that two annihilation quanta react independently in 2.5 cm thick layer is 16 times smaller for the plastic detector than in the detector made of LSO crystals.

**Arguments in favor of polymer scintillators:**
Here we argue that disadvantages due to the low detection efficiency and negligible probability for photoelectric effect can be compensated by large acceptance and improved time resolution achievable with polymer scintillator detectors.

a) **Compton effect:**

In polymer scintillators probability for the photoelectric effect is indeed negligible. But still it is possible to use events related to Compton effect inside the detector. The maximum energy deposition of electrons from the Compton edge is equal to about 340 keV. In Fig. 1 we show Compton scattered electron energy distributions for (i) the energy of gamma quanta reaching the detector without scattering in the patient's body, (ii) after the scattering through an angle of 30 degrees and (iii) after scattering under an angle of 60 degrees. The presented distributions show that in order to limit registration of quanta scattered in the patient to the range from 0 to 60 degrees (as used in the currently produced tomographs) one has to use an energy threshold of about 200 keV.

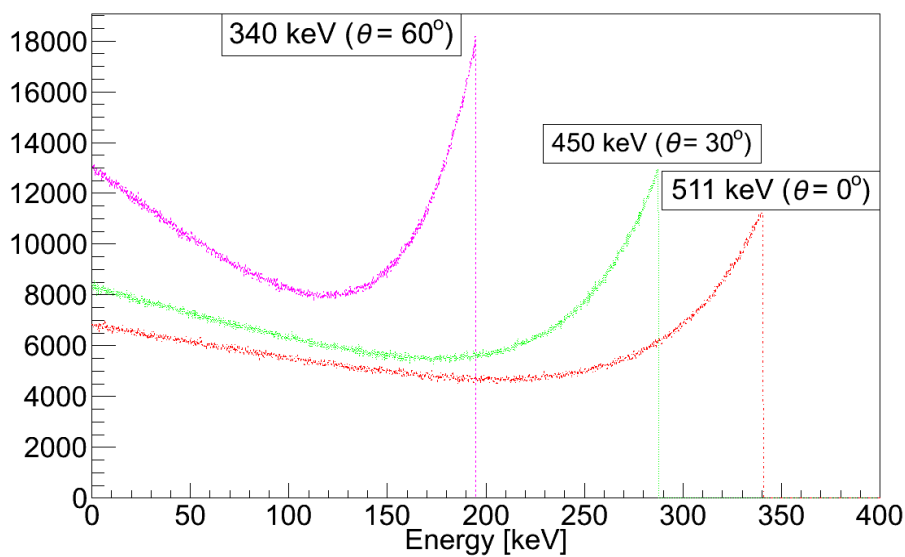

**Figure 1.** Energy distribution of electrons scattered in the Compton effect by gamma quanta with an energy shown in the plot [9]. The distributions were made without taking into account the energy resolution, which for plastic detectors is about 18% (compared to energy resolution of about 12% for LSO blocks [3]).

b) **Acceptance:**

A low efficiency may be compensated when using: (i) a large diagnostic chamber, (ii) 3D mode for reconstruction, and (iii) ticker layers of scintillators. For this, however, novel detector solutions are required which would enable (i) to build large chambers in a cost effective way, (ii) to decrease the noise for the effective application of the 3D image reconstruction, and (iii) to reconstruct a depth of interaction of gamma quanta in the thick detector material.

The above requirements may be realized when using polymer scintillators which are much cheaper than crystals. Typically, for plastic scintillators, price per unit of volume is more than a factor of 30 lower than for BGO crystals and more than a factor of 80 lower than for LSO. In addition plastic scintillators can be easily produced in large sizes and various shapes, contrary to expensive inorganic crystals. In the 3D mode the geometric acceptance of e.g. one meter long chamber would increase on average by a factor of about five in the comparison to the present PET detectors. This feature in combination with the five times larger longitudinal field of view would lead to about 25 times more effective registration of pairs of annihilation quanta. Large acceptance would compensate the smaller efficiency, which in addition can also be increased by tick or multiple scintillator layers. Additionally: a larger longitudinal field of view would allow for simultaneous imaging of larger fraction of the body. In the case of current PET scanners such image of a whole body requires performance of many independent measurements in steps taken moving the patient inside the tomograph by about half the width of the ring [5,3]. Thus, in case of the whole body examination, an increase of the longitudinal field view by a given factor would increase statistics of registered events by the same amount.

   c) **Time resolution**

The TOF resolution obtainable with plastic scintillators may be even better than 100 ps for large detector (even in the scale of meter [10,11]). The time resolution depends predominantly on the ratio of number of photo-electrons produced in the converter to the duration of the signal. This is best for the plastic scintillators with decay time less than 2 ns and with still large light output of 10 000 photons / MeV. This can be compared to 40 ns decay time of LSO crystals with light output of 32 000. Better time properties will decrease the noise along the line of response and should allow for an effective 3D image reconstruction for the detector with large field of view. It can be shown, that when taking into account TOF in the image reconstruction, the sensitivity (image contrast) increases inversely proportional with the time resolution and directly proportional to the size of the examined object [5]. Absolute values of the improvement can be determined knowing that for an object with a diameter of 40 cm the time resolution of current PET scanners, which is about 600 ps, improve the image contrast by a factor of about four [5,6]. Improving the time resolution down to 100 ps would improve it by another factor of six.

**Novel detector concepts:**

The proposed detector concepts are shown schematically in Fig. 2. Such solutions may enable construction of large diagnostic chambers but they will require large blocks or strips of scintillator. For this purpose plastic scintillators are at most appropriate because their light attenuation length is very large (typically about 2 meters) contrary to rather short (typically 20 cm) light attenuation length of inorganic crystals. In addition plastic scintillators possess relatively low refractive index equal to 1.58 which for long path length implies smaller time spread in comparison to inorganic crystals characterized with larger refractive index as e.g. 1.85 for LSO or 2.15 for BGO.

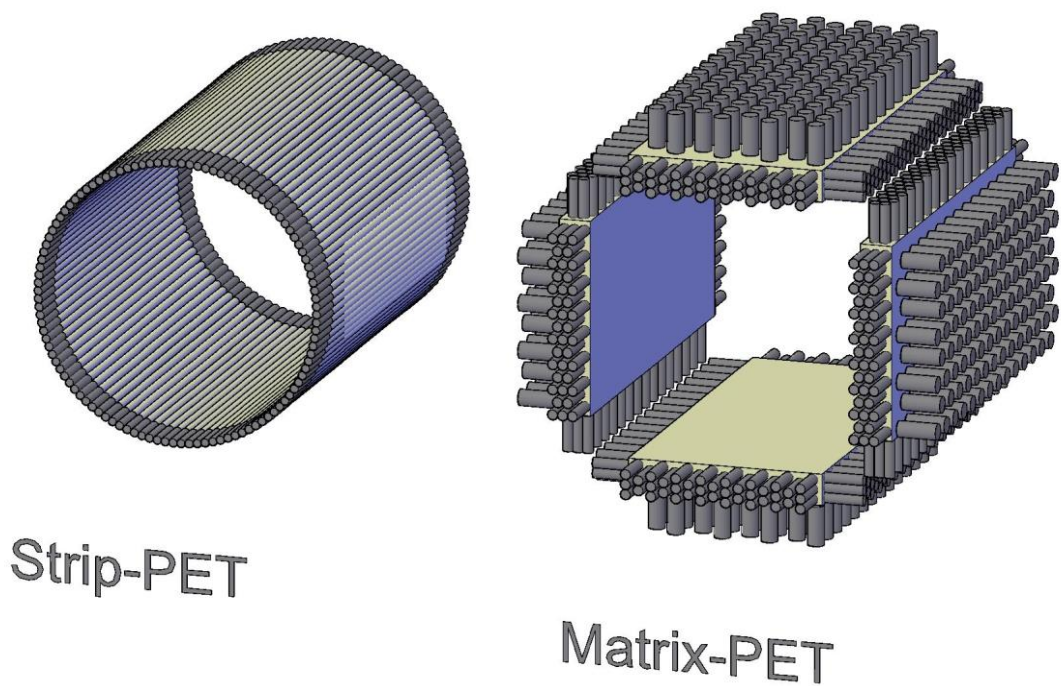

**Figure 2.** Schematic view of one of the possibilities of arrangement of scintillation elements and photomultipliers for the diagnostic chambers of STRIP-PET and MATRIX-PET detector systems.

In this contribution we discuss the solution referred to as MATRIX-PET and the other concept called STRIP-PET [12] is described more detailed in the accompanying paper [13].

**Matrix-PET**
Uniqueness of the Matrix-PET constitutes the solution of light collection allowing for the conversion to the electric signal of direct light [14,15]. The idea is demonstrated schematically in the left panel of Fig. 3. This method allows to achieve a time resolution which is not affected by the deformation of light pulses due to reflections at scintillators surfaces. Such PET detector would consist of organic scintillator plates. The plates could be set in many ways so as to cover the whole body of the patient, for example as it is shown in the right panel of Fig. 2. The measurement of time and amplitude of light signals is carried out by photomultipliers matrix arranged around the chamber. The interaction point within the plane of the plate can be reconstructed based on both: (i) the distribution of time of the signals from photomultipliers and (ii) distribution of amplitudes of the recorded signals. Such solution enables also determination of the depth at which the gamma quantum has been absorbed (DOI) on the basis of the distribution of amplitudes of signals from photomultipliers arranged around the sides of the plates. This feature allows using of thick plates without worsening of spatial resolution due to "the DOI problem" occurring in the current PET tomographs. This solution would also enable effective usage of the TOF method permitting the determination of the annihilation point along the line-of-response based on the time difference in reaching the different scintillation plates by the gamma quanta, as demonstrated in the right panel of Fig. 3. Polymer scintillators allow to obtain the time resolution better than 100 ps compared to 600 ps achievable in a current PET scanners. Such accuracy of TOF determination may significantly improve the sensitivity (image contrast) which increases inversely proportional with the time resolution and directly proportional to the size of the examined object [5].

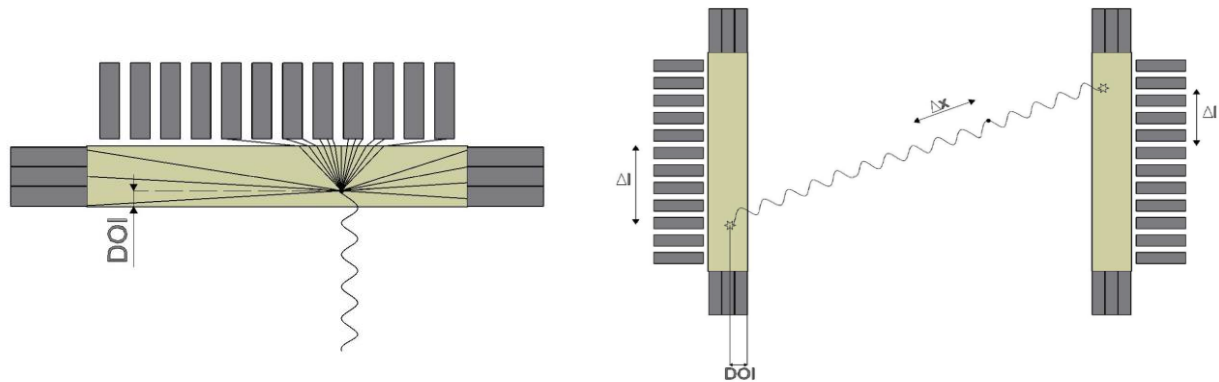

**Figure 3.** (Left) Schematic view of the cross section of a single detection plate for the MATRIX-PET. (Right) The hit position versus the center of the scintillator (Δl) is determined based on time difference measured on both sides of the scintillator plate and independently based on the amplitude distributions of signals in the matrix of photomultipliers surrounding the plate. A position (Δx) along the LOR is determined from time difference measured between two plates.

**Summary:**
In this contribution we have presented novel solution for the detector system which may be applied in the Positron Emission Tomography. We have argued that it may be possible to build a PET modality with large longitudinal field of view based on the polymer scintillators. A low efficiency for detection of annihilation quanta and low probability for photoelectric effect in polymer scintillators can be compensated by large acceptance and very good time resolution achievable with this kind of detectors. Two different designs of the diagnostic chamber have been presented which make use of large blocks of relatively cheap organic scintillators with specially designed redouts and which make use of timing of signals for reconstruction of points where gamma quanta underwent reaction inside the detector material. An especially promising is a possibility of extension of the diagnostic chamber in strip-PET solution which does not entail an increase in the number of photomultipliers. Such feature would decrease the construction costs of PET scanners enabling simultaneous imaging of the physiological processes throughout the whole body of the patient.

**Acknowledgement:**
Authors acknowledge support of the Foundation for Polish Science and the Polish National Center for Development and Research.

**References:**
[1] Buck A K et al., Economic Evaluation of PET and PET/CT in Oncology.
J Nucl Med 2010; 51: 401.
[2] Polish Health Ministry, Narodowy Program Zwalczania Chorób Nowotworowych,
Budowa sieci ośrodków PET, 2005.
[3] Saha G, Basics of PET imaging. Springer New York 2010;
[4] Humm J L et al., From PET detectors to PET scanners.
Eur J Nucl Med Mol Imaging 2003; 30: 1574.
[5] Karp J S et al., Benefit of Time-of-Flight in PET: Experimental and Clinical Results.
J Nucl Med 2008; 49: 462.

[6] Conti M., State of the art and challenges of time-of-flight PET. Physica Medica 2009; 25: 1.
[7] Schaart D R et al., LaBr$_3$:Ce and SiPMs for TOF PET: achieving 100ps coincidence resolving time. Phys Med Biol 2010; 55: N179.
[8] Scintillation Products Data sheet, Saint-Gobain Ceramics & Plastics 2011.
[9] Niedźwiecki Sz, Studies of detection of gamma radiation by means of organic scintillator detectors in view of positron emission tomography. Diploma Thesis 2011.
[10] Sugitate T et al., 100 cm long TOF scintillation counters with rms resolution of 50 ps. Nucl Instr & Meth 1986; A249: 354.
[11] Nishimura S et al., Systematic studies of scintillation detector with timing resolution of 10 ps for heavy ion beam, Nucl Instr & Meth 2003; A510: 377.
[12] Moskal P et al., Novel detector systems for the positron emission tomography, Bio-Algorithms and Med-Systems 2011; 7: 73; arXiv:1305.5187 [physics.med-ph].
[13] Moskal P et al., these proceedings.
[14] Brauksiepe S et al., COSY-11, an internal experimental facility for threshold measurements, Nucl Instr and Meth 1996; A376:397-410.
[15] Anton G et al., AMADEUS: A New type of large area scintillation detector with position, energy and time-of-flight determination, Nucl Instr and Meth 1991; A310: 631-635.